\documentclass[12pt]{article}
\usepackage[utf8]{inputenc}
\usepackage[T2A, T1]{fontenc}

\usepackage{epsfig}
\usepackage[a4paper, width=170mm,top=20mm,bottom=20mm]{geometry}
\usepackage{amsfonts}
\usepackage{amsmath}
\usepackage{amssymb}
\usepackage{enumerate}
\usepackage{multicol}
\usepackage{amsthm}

\theoremstyle{definition}

\title{\textbf{  Chemical Continuous Time Random Walks under Anomalous Diffusion}}
\author{Hong Zhang, Guohua Li}
\date{\today}
\begin{document}
\maketitle
\begin{abstract}
 Chemical master equation plays an important role to describe the time evolution of  homogeneous chemical system. In addition to the reaction process,  it is also accompanied by physical diffusion of the reactants in complex system that is generally not homogeneous, which will result in  non-exponential waiting times for particle reactions and diffusion. In this paper we shall introduce a chemical continuous time random walk under anomalous diffusion model based on renewal process to describe the general reaction-diffusion process in the heterogeneous system, where the waiting times are arbitrary distributed. According to this model, we will develop the systematic stochastic theory including generalizing the chemical diffusion master equation, deriving the corresponding  mass action law, and extending the Gillespie algorithm. As an example, we analyze the monomolecular $A\leftrightarrow B$ reaction-diffusion system for exponential and power-law waiting times respectively, and show the strong fractional memory effect of the concentration of the reactants on the history of the concentration in power-law case.
\end{abstract}
\section{Introduction}  
Anomalous diffusion behaviors have attracted great interest in various
fields of physics, biological, chemical, and environmental sciences. \cite{TB1996,MK2000} The main characteristics of  anomalous diffusion is that the
mean square displacement scales as a nonlinear power law in time, i.e., $<\Delta x^2>\sim t^{\beta}$.
Faster than linear scaling $(\beta>1)$ is corresponding to as superdiffusion, slower than linear scaling $0<\beta<1$ is referred
to as subdiffusion, and when  $\beta=1$ it is normal diffusion.

The continuous time random walk (CTRW) with long-tailed waiting time  has been
often evoked as a suitable description of anomalous diffusion.  In this model, the particle begins its jump at
time $t = 0$ and then traps in a position for a random waiting time until it jumps away.
In recent years there has been considerable effort to investigate the chemical reactions (e.g., reversible or irreversible conversion to a
different species \cite{SSS2006,LHW2008,SSS2007},
spontaneous evanescence \cite{AYL2010}, fluorescence recovery after photobleaching \cite{YAL2014}, or linear reaction dynamics \cite{HLW2006}, etc.) under anomalous diffusion using different continuous time random walks models. In 2010,  Fedotov considered three CTRW models: model $A, B, C$ with nonlinear reactions under varying assumptions for the effect of reaction on random waiting time \cite{F2010}. In 2013, Angstmann et al. generalized the model $B$
where it is assumed that the created particles will draw  a new  waiting time to be in space- and time-dependent force fields, and applied this generalized model with space-dependent traps to obtain a fractional Fokker-Plank equation (FFPE) with space-dependent anomalous exponential for an ensemble of particles \cite{ADH2013}.
However, for the complex reaction-diffusion system, such as the system including more than three reactions each of which has many reactants and products performing anomalous diffusion, the CTRW approach is out of work due to the technical difficulty that it is quite hard for the balance equation to remember all the effects of  so many complex reactions on the probability evolution for every particle and of coupling relations between the reactions and anomalous diffusion.

Fortunately, there is a chemical master equation  (CME) which can provide a stochastic approach to capture the time behavior of a spatially homogeneous chemical
system in particle number space.  The advantage of the CME is that it can deal with the system including many reactions,
and has a firmer physical basis than the deterministic  reaction-rate equations. Besides, the Gillespie algorithm based on the spatially homogeneous master
equation  is proved to be straightforward to
implement on a digital computer. \cite{G1977}
Recently a generalized chemical diffusion master equation (CDME) is developed to describe
the normal diffusion
process for a given n-particle probability density coupled by the reaction dynamics similar in form to a chemical master equation by using
creation and annihilation operators.\cite{RF2022}
But up to now it is still a challenge to  involve anomalous diffusion into CME.

On the other side, the system that both CME and CDME consider is the well-mixed  chemical reaction system in which each reaction is generated  by Poisson process, that corresponds to the exponential distributed waiting time. \cite{GQ2017,RW2023}
However, in various real-world systems the inter-reaction times typically obey non-Poissonian distribution. \cite{MR2018}
In particular, inter-reaction times typically obey long-tailed distributions. Examples of long-tailed distributions of inter-reaction
times include population dynamics, epidemic processes,
finance, and so on. The non-Poissonian chemical process defines
an inhomogeneous chemical continuous time random walk (CCTW) in particle number space. \cite{AD2017} Based on the CCTW Zhang et. al.  proposed a  stationary generalized chemical master equation which can
model intracellular non-Markovian intracellular processes with molecular memory. [17-20]

In this paper we will consider the complex heterogeneous system involving both the arbitrary non-Poisson chemical process  and the anomalous diffusion process of the particles. We shall overcome the technical difficulties to develop a systematic stochastic approach for the generalized CDME, the generalized mass law action, and the generalized Gillespie algorithm based on  renewal processes used in CCTW, where inter-event times are independently generated
from a given distribution \cite{MR2018,C1962,F1971}, and on the continuous time random walk model, where it is assumed that the reactants can move from one space to the other, and at each space or site the reactants can react according to the chemical reaction law, but do not react in the process of jumping.
Moreover, we will take the anomalous monomolecular $A\leftrightarrow B$ reaction-diffusion system as an example, and obtain the corresponding mass action law for exponential and power-law distributed waiting times respectively, and show the strong fractional memory effect of the concentration of the reactants on the history of the concentration in the case of power-law distribution.

\section{The renewal process}
We assume that there are $\gamma$
different species that participate in $m$ different reactions and can diffuse in $N$ positions. The chemical species are denoted by $S_j$, where
$j = 1,...,\gamma$; the $\upsilon$ positions are denoted by $x,y,z,...$, the corresponding particle numbers at $x$ at time $t$ are
denoted by $k_{xj}$.  The state vector of particle numbers is a random $\upsilon\gamma$-dimensional vector
$k=(k_{x1},k_{x2},...,k_{x\gamma},...,k_{z1},k_{z2},...,k_{z\gamma})^{T}$,
where the superscript $T$ denotes the
transpose.
A single event of reaction and diffusion are characterized  by the reaction and diffusion waiting times.
We denote the probability density function (PDF) for the waiting time $\tau_{ix}$ of reaction $i$ at $x$ by $p_{ix}(t)$ and denote the PDF of the waiting time $\tau_{dlx}$ for the particle $l$ diffusing away from $x$  by $p_{dlx}(t)$, respectively.

We now consider a special renewal process where
$N(t)=\sup\{{n\in N: T_n\leq t}\}$ is the number of renewal in $[0,t]$ with $N(0)=0$ if $X_1<t$, where $T_n$ is the time after $n$ renewal step and is the sum $T_n=\tau_1+\tau_2+...+\tau_n$, $n\geq 1$ where $\tau_1, \tau_2,...,\tau_{n}$ are a sequence of nonnegative independent distributions random variables. For each $\tau_n, n\geq 1$ is the waiting time for the $n-th$ update. It is assumed to be
the minimum time of all waiting times of reactions and diffusions, that is, $\tau_n=\min\{\tau_{ix},\tau_{dlx}: i=1,2,...,m, l=1,2,...,\gamma, x=1,2,...,\upsilon\}$.

One can find $N(t)=N_r(t)+N_d(t)$ where $N_r(t)$ is the number of reaction renewals in $[0,t]$, and $N_d(t)$ is the number of diffusion renewals in $[0,t]$.
The state vector will change for each reaction renewal and each diffusion renewal.
During a reaction $i$ at $x$ the loss (gain) in particle
number $k_{xj}$  is denoted by $r_{ixj}\in N$ ($p_{ixj}\in N$). These coefficients are typically, but need not be, given by the law of
mass action. Thus, the impact of reaction i on the state
space can be expressed as \cite{AD2017}
$$\sum_{j}r_{ixj}S_j\rightarrow \sum_{j}p_{ixj}S_j$$
The stoichiometric coefficients $s_{ixj} = p_{ixj}-r_{ixj}$ denote the
net change in each species $j$ due to each reaction $i$ at space $x$, and form the reaction transforming vector for the $n-th$ step $s_{ix}=(0,...,0,s_{ix1},s_{ix2},...,s_{ix\gamma},0,...,0)$.
For each particle $j$, its space at time $t$ is $Z_j(t)=Z_j(0)+\sum_{n=1}^{N_{d_j}}\xi_n$ where $\xi_n$ is the jumping length for the $n-th$ jumping  step of particle $j$, $N_{d_j}(t)$ is the number of diffusion renewals of particle $j$ in $[0,t]$ and the relation with $N_d(t)$ is $\sum_{j}N_{d_j}(t)=N_{d}(t)$. If in one step an $i$ reactant moves from $x$ to $y$, then after this step the number $k_{xl}$ of the reactant $l$ reduces $1$ at $x$  and $k_{yl}$ adds $1$ at $y$, so we can find  the
diffusion transforming vector $s_{dlxy}=\binom{0,...,-1,0,...,1,0,....0}{      xl,......,yl}$ (all except $(x-1)\gamma +l$-th and $(y-1)\gamma +l$-th components are $0$).

\subsection{Distribution of the renewal waiting time}

We now investigate the distribution of the renewal waiting time $\tau_n=\min\{\tau_{ix},\tau_{dlx}: i=1,2,...,m, l=1,2,...,\gamma, x=1,2,...,\upsilon\}$. We find
\begin{eqnarray}
   P(t_n\leq t)&=&1-P(\min\{\tau_{ix},\tau_{dlx}:i=1,2,...,m, l=1,2,...,\gamma, x=1,2,...,\upsilon\}\geq t\}) \nonumber\\
    &=& 1-\Pi_{i=1,2,...,m, l=1,2,...,\gamma, x=1,2,...,\upsilon} \Psi_{ix}^{r}(t,k)\Psi_{lx}^{d}(t,k).
    \label{eq:waitingtime1}
\end{eqnarray}
  In the second equation we used the property of independence of all waiting times.
Here,  $\Psi_{ix}^{r}(t,k)P(\tau_{ix}\geq t)$ and $\Psi_{lx}^{d}(t,k)=P(\tau_{dlx}\geq t)$.

One can easily see that the survival probability $\Phi(t,k)=P(\tau_n \geq t)$ of the event that there is no diffusion and no reaction in the system in time interval $[0, t]$ is $$\Phi(t,k)=\Pi_{x=1,2,...,\upsilon;i=1,2,...,m}\Psi_{ix}^{r}(t,k)\Pi_{x=1,2,...,\upsilon;l=1,2,...,\gamma}\Psi_{lx}^{d}(t,k).$$

To get the exact distribution $P(t_n\leq t)$ of $\tau_n$ in Eq. \eqref{eq:waitingtime1}, we first discuss $\Psi_{ix}^{r}(t,k)=P(\tau_{ix}\geq t)$, namely, the survival probability for reaction $i$ not occurring at $x$ in the time interval $[0,t]$. Note that appearance of the reaction of $i$ at $x$  means that at least one group of reaction $i$ at $x$  reacts. We assume that there are $h_{ix}(k)$  groups  of reactants for reaction $i$ at space $x$. Let $\tau_{ixe}( e=1,2,...,h_{ix}(k))$ be the waiting times which are assumed to be i.i.d.(independent,identically distributed).
One notes that $\tau_{ix}=\min\{\tau_{ixe}:e=1,2,...,h_{ix}(k)\}$, and then
 \begin{eqnarray}
   P(\tau_{ix}\geq t)&=&P(\min\{\tau_{ixe}:e=1,2,...,h_{ix}(k)\}\geq t\})= \prod_{e=1}^{h_{ix}} P( \tau_{ixe}\geq t),
   \label{eq:probability}
\end{eqnarray}
where we used  the fact that the waiting times for all groups of reactants for reaction $i$  are independent.

Let $p_{ix}(t)$ be the waiting time PDF (probability density function) for $\tau_{ixe} (e=1,2,...,h_{ix}(k))$, then from Eq. \eqref{eq:probability} we find
\begin{eqnarray}
   P(\tau_{ix}\geq t)&=&(\int_{t}^{+\infty}p_{ix}(t')dt')^{h_{ix}(k)}.
   \label{eq:notreaction}
\end{eqnarray}

Therefore, the distribution of $\tau_{ix}$ is $P(\tau_{ix}\leq t)=1-(\int_{t}^{+\infty}p_{ix}(t')dt')^{h_{ix}(k)}$, and
the PDF $\psi_{ix}^{r}(t,k)$ for $\tau_{ix}$, which is the PDF for at least one group of reactants for reaction $i$ at $x$ first occurring, can be got by taking the differentiation of $P(\tau_{ix}\leq t)$ with respect to $t$, that is,
\begin{eqnarray}
\psi_{ix}^{r}(t,k)=h_{ix}(k)p_{ix}(t)(\int_{t}^{\infty}p_{ix}(t')dt')^{h_{ix}(k)-1}.
\label{eq:pdfforreactionwaitingtime1}
\end{eqnarray}

Note that $\psi_{ix}^{r}(t,k)$ can also be obtained by the other way as following.

Firstly, one finds
\begin{eqnarray}
   P(\tau_{ix}\geq t)&=&P(\min\{\tau_{ixe}:e=1,2,...,h_{ix}(k)\}\geq t\}) \nonumber\\
   &=& \sum_{e=1}^{h_{ix}}(k) P(\tau_{ix}=\tau_{ixe}, \tau_{ixe}\geq t)\nonumber\\
       &=& h_{ix}(k) P(\tau_{ix}=\tau_{ixe}, \tau_{ixe}\geq t)\nonumber\\
         &=& h_{ix}(k) P(\tau_{ixf}\geq\tau_{ixe}, \tau_{ixe}\geq t: f=1,2,...,h_{ix}(k), f\neq e)\nonumber\\
&=&h_{ix}(k) P(\tau_{ixf}\geq\tau_{ixe}, \tau_{ixe}\geq t)\nonumber\\
   &=&h_{ix}(k)\int_{t}^{+\infty}p_{ix}(t')dt'(\int_{t'}^{+\infty}p_{ix}(t'')dt'')^{h_{ix}(k)-1}.
\end{eqnarray}
where we used the property of independence of multiple random variables and the fact that the waiting times for all groups of reaction $i$  are independent.

Therefore, the PDF $\psi_{ix}^{r}(t,k)$ for $\tau_{ix}$ can be got by taking the differentiation of the distribution of $\tau_{ix}$ (i.e., $P(\tau_{ix}\leq t)=1-P(\tau_{ix}\geq t)$ ) with respect to $t$, that is,
\begin{eqnarray}
\psi_{ix}^{r}(t,k)=h_{ix}(k)p_{ix}(t)(\int_{t}^{\infty}p_{ix}(t')dt')^{h_{ix}(k)-1},
\end{eqnarray}

Analogously, we can discuss $\Psi_{lx}^{d}(t,k)=P(\tau_{dlx}\geq t)$ in Eq. \eqref{eq:waitingtime1}, namely, the survival probability for all $l$ reactants not diffusing away from $x$ in time interval $[0,t]$.  We denote the waiting time for one $l$ diffusing away from $x$ by $\tau_{dlxf}(f=1,2,...,k_{xl})$, and assume that the waiting times for all $l$ reactants are i.i.d. Let $p_{dlx}(t)$ be the waiting time PDF  for $\tau_{dlxf}$.
One notes $\tau_{dlx}=\min\{\tau_{dlxf}: f=1,2,...,k_{xl}\}$, and then
 \begin{eqnarray}
   P(\tau_{dlx}\geq t)&=&P(\min\{\tau_{dlxf}:f=1,2,...,k_{xl}\}\geq t\})=(\int_{t}^{+\infty}p_{dlx}(t')dt')^{k_{xl}}.
   \label{eq:notdiffusion}
\end{eqnarray}

The distribution of $\tau_{dlx}$ is $P(\tau_{dlx}\leq t)=1-(\int_{t}^{+\infty}p_{dlx}(t')dt')^{k_{xl}}$,
the PDF $\psi_{lx}^{d}(t,k)$ of $\tau_{dlx}$ can be got by taking the differentiation of $P(\tau_{dlx}\leq t)$ with respect to $t$ as
\begin{eqnarray}
\psi_{lx}^{d}(t,k)=k_{xl}(k)p_{dlx}(t)(\int_{t}^{\infty}p_{dlx}(t')dt')^{k_{xl}-1}.
\end{eqnarray}

Now the exact expression of Eq.\eqref{eq:waitingtime1} can be easily obtained from Eqs.\eqref{eq:notreaction} and \eqref{eq:notdiffusion}. Next, we will find the other representation of this distribution \eqref{eq:waitingtime1}. Firstly, we discuss the event that after waiting for $t$, the reaction $i$ at space $x$ in the system first occurs but none of other reactions  and none of diffusions has happened, which implies that the waiting time of this reaction in this renewal step is the minimum waiting time. The distribution for this event can be described by
\begin{eqnarray}
   P(\tau_{n}=\tau_{ix}, \tau_{n}\leq t)=P(\tau_{jx}\geq\tau_{ix}, \tau_{dlx}\geq\tau_{ix}, \tau_{ix}\leq t),
\end{eqnarray}
for $j=1,2,...,m$ and $j\neq i$.

By using the property of independence of multiple random variables, one obtains
\begin{eqnarray}
  P(\tau_{n}=\tau_{ix}, \tau_{ix}\leq t) &=&\int_{0}^{t}\psi_{ix}^{r}(t')dt'(\int_{t'}^{+\infty}\psi_{jx}^{r}
  (\tau)d\tau,....,\int_{t'}^{+\infty}\psi_{lx}^{d}(\tau)d\tau)\nonumber\\
    &=&\int_{0}^{t}\psi_{ix}^{r}(t')[\Pi_{j\neq i}\Psi_{jx}^{r}(t',k)\Pi_{y\neq x;i=1,2,...,m}\Psi_{iy}^{r}(t',k)]dt'.
\end{eqnarray}

Therefore, we can obtain the PDF for the event that after waiting for $t$, the reaction $i$ at space $x$ first occurs but none of other reactions  and  none of diffusions has happened  by differential with respect to $t$ of $P(\tau_{n}=\tau_{ix}, \tau_{ix}\leq t)$, that is,
\begin{eqnarray}
\phi_{ix}^{r}(t,k)&=&\frac{\partial P(\tau_{n}=\tau_{ix}, \tau_{ix}\leq t)}{\partial t}\nonumber\\
&=&\psi_{ix}^{r}(t)\Pi_{j\neq i}\Psi_{jx}^{r}(t,k)\Pi_{y\neq x;i=1,2,...,m}\Psi_{iy}^{r}(t,k)\Pi_{x=1,2,...,\upsilon;l=1,2,...,\gamma}\Psi_{lx}^{d}(t,k).
\end{eqnarray}
 When the diffusion is not considered, and the chemical process focuses one space $x$, $\phi_{ix}^{r}(t,k)$ becomes $\phi_{ix}^{r}(t,k)=\psi_{ix}^{r}(t)\Pi_{j\neq i}\Psi_{jx}^{r}(t,k)$, which is the result in the non-Markovian case in Refs.\cite{ZZ2019,AD2017} obtained from other way.

Analogously, if $\tau_n$ is the waiting time $\tau_{dlx}$ for the diffusion of the reactant $l$ from $x$, then we find the PDF for the event that the diffusion of the reactant $l$ from $x$ first occurs but none of other diffusions  and  none of reactions has happened by differential with respect to $t$ of $P(\tau_{n}=\tau_{dlx}, \tau_{dlx}\leq t)$ as
\begin{eqnarray}
\phi_{lx}^{d}(t,k)&=&\psi_{lx}^{d}(t,k)\Pi_{l'\neq l}\Psi_{l'x}^{d}(t,k)
\Pi_{y\neq x;l=1,2,...,\gamma }\Psi_{ly}^{d}(t,k)\nonumber\\
&&\cdot\Pi_{i=1,2,...,m;x=1,2,...,\upsilon}\Psi_{ix}^{r}(t,k).
\end{eqnarray}

Finally, we find
\begin{eqnarray}
-\frac{\partial\Phi(t,k)}{\partial t}&=&\sum_{x=1}^{\upsilon}\sum_{i=1}^{m}\phi_{ix}^{r}(t,k)+\sum_{x=1}^{\upsilon}\sum_{l=1}^{\gamma}\phi_{lx}^{d}(t,k),
\end{eqnarray}
which means
\begin{eqnarray}
\Phi(t,k)&=&\int_{t}^{\infty}\sum_{x=1}^{\upsilon}\sum_{i=1}^{m}\phi_{ix}^{r}(t',k)+\sum_{x=1}^{\upsilon}\sum_{l=1}^{\gamma}\phi_{lx}^{d}(t',k)dt',~~
\end{eqnarray}
and then
\begin{eqnarray}
\Phi(u,k)&=&\frac{1}{u}[1-(\sum_{x=1}^{\upsilon}\sum_{i=1}^{m}\phi_{ix}^{r}(u,k)+\sum_{x=1}^{\upsilon}\sum_{l=1}^{\gamma}\phi_{lx}^{d}(u,k))],
\label{eq:expression2relation}
\end{eqnarray}
where $\Phi(u,k)$,  $\phi_{ix}^{r}(u,k)$ and $\phi_{xl}^{d}(u,k)$ are the Laplace $t\rightarrow u$ transform of
$\Phi(t,k)$,  $\phi_{ix}^{r}(t,k)$ and $\phi_{xl}^{d}(t,k)$, respectively.

\section{The chemical master equation under anomalous diffusion in one-dimensional finite lattice }

We then want to investigate the evolution of the state vector $k$ at time $t$ in this renewal process, that is, to obtain the master equation of the chemical continuous time random walks under anomalous diffusion model (see Figure.1).

\begin{figure}
	\begin{center}
		\includegraphics[width=10.8cm]{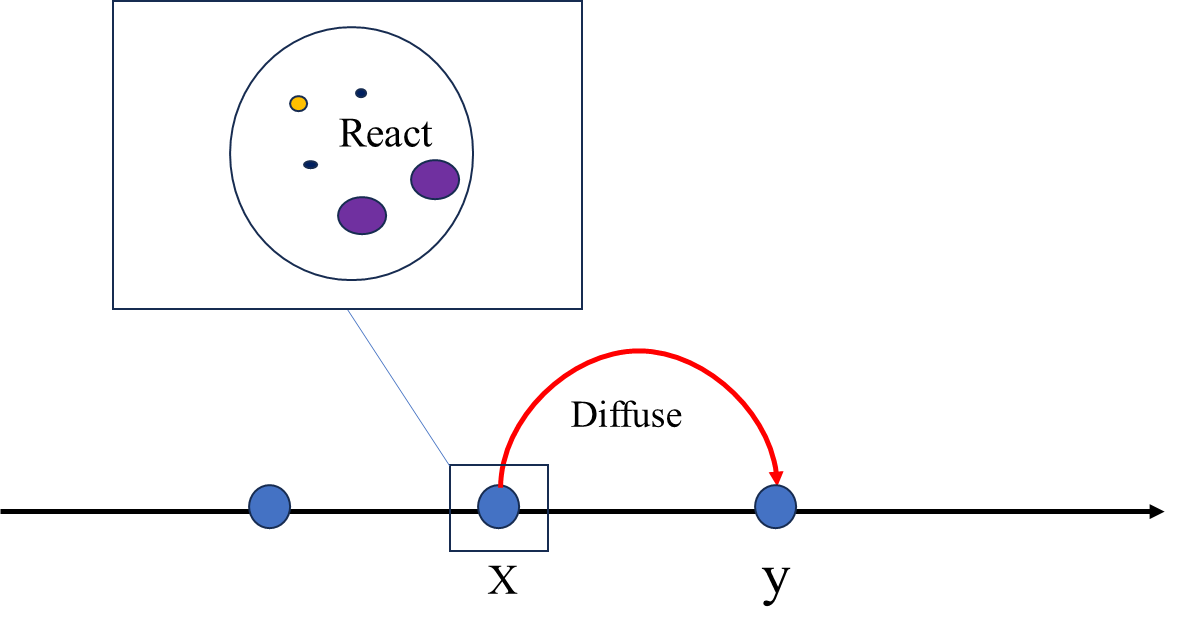}
	\end{center}
	\caption{The chemical continuous time random walks under anomalous diffusion model where the reactants wait some random time with arbitrary distribution at a space and then diffuse  to other space with certain probability,  and meanwhile,  the reactants react according to chemical law at each space.
		\label{snapshots}}
\end{figure}

Let $P(k, t)$ denote the probability
distribution $Prob[k(t)=k]$.
Firstly, according to the renewal theory,
the probability for the system at state $k$ at $t$ equals to the probability for the system just arriving at $k$ at earlier time $t'\leq t$, and not changing its state until $t$,
so we get the balance equation
\begin{eqnarray}
P(k,t)=\int_{0}^{t}dt'\sum_{n=0}^{\infty}R_n(k,t')\Phi(t-t',k),
\label{eq:renewalrelation}
\end{eqnarray}
where $R_{n}(k,t)$ is the joint density of just arriving at $k$ at time $t$ after $n$ steps, and satisfying
\begin{eqnarray}
R_{n+1}(k,u)&=&\int_{0}^{t}\bigg\{\sum_{i=1}^{m}\sum_{x=1}^{\upsilon}R_n(k-s_{ix},t')\phi_{ix}^{r}(t-t',k-s_{ix})
\nonumber\\
&&+\sum_{l=1}^{\gamma}\sum_{x=1}^{\upsilon}\sum_{y\neq x}[R_n(k-s_{dlxy},t')\phi_{lx}^{d}(t-t',k-s_{dlxy})\lambda_{xy}]\bigg\}dt',
\label{eq:renewalbalanceequation}
\end{eqnarray}
for nonnegative integer $n$,
and
\begin{eqnarray}
R_{0}(k,t)=P(k,0)\delta(t).
\end{eqnarray}
Here, the symbol $\lambda_{xy}$ represents the transition probability for the reactant jumping from $x$ to $y$ and $P(k,0)$ is the initial distribution. Notice that
Eq.\eqref{eq:renewalbalanceequation} is a balance equation, too.  It bases on the fact that the event for the system just arriving at $k$ at earlier time $t'\leq t$ after $n+1$ step is equivalent to the event for the system just arriving at $k-s_{ix}$ at earlier time $t'\leq t$ after $n$ step (that is,  $T_{n}=t'$), and waiting for $t-t'$ to take the $n+1-$th step (i.e., reacting) at $t$ (that is, $T_{n+1}=t$) and meanwhile changing its state from $k-s_{ix}$ to $k$, or for the system just arriving at $k-s_{dlxy}$ at $t'\leq t$ after $n$ step, and waiting for $t-t'$ to take the $n+1-$th step (i.e., diffusing) at $t$ and meanwhile changing its state from $k-s_{dlxy}$ to $k$.

By using the Laplace $t\rightarrow u$ transform of Eq.\eqref{eq:renewalbalanceequation}, we get
\begin{eqnarray}
R_{n+1}(k,u)&=&\sum_{i=1}^{m}\sum_{x=1}^{\upsilon}R_n(k-s_{dlxy},u)\phi_{ix}^{r}(u,k-s_{ix})
\nonumber\\
&&+\sum_{l=1}^{\gamma}\sum_{x=1}^{\upsilon}\sum_{y\neq x}[R_n(k-s_{dlxy},u)\phi_{lx}^{d}(u,k-s_{dlxy})\lambda_{xy}],
\label{eq:renewalbalanceequationlaplace}
\end{eqnarray}
Here, $R_{n}(k,u)$ is the Laplace $t\to u$ transform  of $R_{n}(k,t)$. The first balance equation \eqref{eq:renewalrelation} tells that the probability for the system at state $k$ at $t$ is

Let $R(k,u)=\sum_{n=0}^{\infty} R_n(k,u)$. Then
\begin{eqnarray}
R(k,u)&=&\sum_{i=1}^{m}\sum_{x=1}^{\upsilon}R(k-s_{ix},u)\phi_{ix}^{r}(u,k-s_{ix})\nonumber\\
&&+\sum_{l=1}^{\gamma}\sum_{x=1}^{\upsilon}\sum_{y\neq x}[R(k-s_{dlxy},u)\phi_{lx}^{d}(u,k-s_{dlxy})\lambda_{xy}]+P(k,0).
\label{eq:renewrelation2}
\end{eqnarray}

Taking the Laplace transform of Eq.\eqref{eq:renewalrelation},  we get
\begin{eqnarray}
P(k,u)=\sum_{n=0}^{\infty}R_n(k,u)\Phi(u,k)
\label{eq:renewrelation3}
\end{eqnarray}
Combining Eqs.\eqref{eq:renewrelation3}, \eqref{eq:renewrelation2} and \eqref{eq:expression2relation} yields
\begin{eqnarray}
P(k,u)&=&\sum_{n=0}^{\infty}R_n(k,u)\Phi(u,k)=R(k,u)\Phi(u,k)\nonumber\\
&&=R(k,u)\frac{1}{u}[1-(\sum_{x=1}^{\upsilon}\sum_{i=1}^{m}\phi_{ix}^{r}(u,k)+\sum_{x=1}^{\upsilon}\sum_{l=1}^{\gamma}\phi_{xl}^{d}(u,k))],
\label{eq:renewalrelation4}
\end{eqnarray}
and then
\begin{eqnarray}
R(k,u)=P(k,u)\frac{1}{\Phi(u,k)},
\label{eq:renewalrelation5}
\end{eqnarray}
where $R(k,u)$ and $P(k,u)$  are the Laplace transforms of $R(k,t)$ and $P(k,t)$, respectively.

From Eq.\eqref{eq:renewalrelation4} we find
\begin{eqnarray}
uP(k,u)&=&R(k,u)-R(k,u)[\sum_{x=1}^{\upsilon}\sum_{i=1}^{m}\phi_{ix}^{r}(u,k)+\sum_{x=1}^{\upsilon}\sum_{l=1}^{\gamma}\phi_{lx}^{d}(u,k)]
\label{eq:renewalrelation6}
\end{eqnarray}
Substituting Eqs.\eqref{eq:renewrelation2} and \eqref{eq:renewalrelation5} into the above equation \eqref{eq:renewalrelation6}, and taking some algebraic operations, we finally obtain the master equation for the chemical continuous time random walks with anomalous diffusion in Laplace space
\begin{eqnarray}
uP(k,u)-P(k,0)&=&\sum_{i=1}^{m}\sum_{x=1}^{\upsilon}P(k-s_{ix},u)\frac{\phi_{ix}^{r}(u,k-s_{ix})}{\Phi(u,k-s_{ix})}
\nonumber\\
&&+\sum_{l=1}^{\gamma}\sum_{x=1}^{\upsilon}\sum_{y\neq x}[P(k-s_{dlxy},u)\frac{\phi_{lx}^{d}(u,k-s_{dlxy})}{\Phi(u,k-s_{dlxy})}\lambda_{xy}]\nonumber\\
&&-\sum_{x=1}^{\upsilon}\sum_{i=1}^{m}P(k,u)\frac{\phi_{ix}^{r}(u,k)}{\Phi(u,k)}-\sum_{l=1}^{\gamma}\sum_{x=1}^{\upsilon}P(k,u)\frac{\phi_{xl}^{d}(u,k)}{\Phi(u,k)}.
\end{eqnarray}
Note that the first item on the right side is just the gain flux and the last two items is the loss flux.
Inverting it into time space we obtain the master equation in time space as following
\begin{eqnarray}
\frac{\partial P(k,t)}{\partial t}&&=\sum_{i=1}^{m}\sum_{x=1}^{\upsilon}\int_{0}^{t}P(k-s_{ix},t')\Theta_{ix}^{r}(t-t',k-s_{ix})dt'
\nonumber\\
&&+\sum_{l=1}^{\gamma}\sum_{x=1}^{\upsilon}\sum_{y\neq x}\int_{0}^{t}[P(k-s_{dlxy},t')\Theta_{lx}^{d}(t-t',k-s_{dlxy})\lambda_{xy}]dt'\nonumber\\
&&-\sum_{x=1}^{\upsilon}\sum_{i=1}^{m}\int_{0}^{t}P(k,t')\Theta_{ix}^{r}(t-t',k)dt'-\sum_{l=1}^{\gamma}\sum_{x=1}^{\upsilon}\int_{0}^{t}P(k,t')\Theta_{xl}^{d}(t-t',k)dt',
\label{master}
\end{eqnarray}
where $\Theta_{ix}^{r}(t,k)$ and $\Theta_{lx}^{d}(t,k)$ are respectively the inverse Laplace transforms $u\rightarrow t$ of
$\Theta_{ix}^{r}(u,k)=\frac{\phi_{ix}^{r}(u,k)}{\Phi(u,k)},$
and
$\Theta_{lx}^{d}(u,k)=\frac{\phi_{xl}^{d}(u,k)}{\Phi(u,k)}.$
Note that when the diffusion process disappears, the above master equation is just the CME for the original non-markovian reaction
system in \cite{ZZ2019, AD2017}, and  $\Theta_{ix}^{r}(t,k)$ is just the memory kernel \cite{ZZ2019}.
Note also that in the steady state we get
\begin{eqnarray}
&&\sum_{i=1}^{m}\sum_{x=1}^{\upsilon}\int_{0}^{t}P(k-s_{ix},t')\Theta_{ix}^{r}(t-t',k-s_{ix})dt'
\nonumber\\
&&+\sum_{l=1}^{\gamma}\sum_{x=1}^{\upsilon}\sum_{y\neq x}\int_{0}^{t}[P(k-s_{dlxy},t')\Theta_{lx}^{d}(t-t',k-s_{dlxy})\lambda_{xy}]dt'\nonumber\\
&&=\sum_{x=1}^{\upsilon}\sum_{i=1}^{m}\int_{0}^{t}P(k,t')\Theta_{ix}^{r}(t-t',k)dt'+\sum_{l=1}^{\gamma}\sum_{x=1}^{\upsilon}\int_{0}^{t}P(k,t')\Theta_{lx}^{d}(t-t',k)dt'.
\end{eqnarray}

\section{Mass action law under anomalous diffusion in one-dimensional finite lattice}
We will  consider the rate law  to describe the macroscopic behavior of reaction-diffusion system. We assume that the whole number $\sum_{x=1}^{\upsilon}\sum_{l=1}^{\gamma}k_{xl}$ of the particles in this system is an invariable constant $n_0$.
Let $n_{xl}$ represent the number for the particle $l$ at position $x$,
and let $P_l(c,x,t)$ represent the probability of the concentration $c=\frac{n_{xl}}{n_0}$ for  $l$ at  $x$ at time $t$. Then  we find $P_l(c,x,t)=\sum_{k: k_{xl}=n_{xl}}  P(k,t)$ which satisfies $\int_0^1P_l(c,x,t)dc=\sum_k  P(k,t)=1$.  Thus,
\begin{eqnarray}
&&\frac{\partial P_l(c,x,t)}{\partial t }=\sum_{k:k_{xl}=n_{xl}}\frac{\partial P(k,t)}{\partial t}.
\end{eqnarray}

Combining with Eq.(\ref{master}), we obtain
\begin{eqnarray}
\frac{\partial P_l(c,x,t)}{\partial t }
&&=\sum_{i:s_{ixl}\neq 0} \sum_{k:k_{xl}=n_{xl}} \bigg[P(k-S_{ix},t') \nonumber\\
&&\cdot\Theta^{r}_{ix}(t-t',k-S_{ix})-P(k,t') \Theta^{r}_{ix}(t-t',k)\bigg] \nonumber\\
&&+\sum_{y\neq x} \sum_{k:k_{xl}=n_{xl}}\bigg[P(k-S_{dlxy},t')\Theta^{d}_{lx}(t-t',k-S_{dlxy})\nonumber\\
&&\cdot\lambda_{xy}+P(k-S_{dlyx},t') \Theta^{d}_{ly}(t-t',k-S_{dlyx})\lambda(x-y)\nonumber\\
&&-P(k,t')\Theta^{d}_{lx}(t-t',k)\lambda_{xy}-P(k,t')\Theta^{d}_{ly}(t-t',k)\lambda_{yx}\bigg].
\end{eqnarray}
Let $a_{ixl}=\frac{s_{ixl}}{n_0}$. Then we get
\begin{eqnarray}
\frac{\partial P_l(c,x,t)}{\partial t }
&&=\int_0^t\bigg\{\sum_{i:a_{ixl}\neq 0}  \bigg[P_l(c-a_{ixl},x,t') \nonumber\\
&&\cdot\Theta^{r}_{ix}(t-t',c-a_{ixl})-P_l(c,x,t') \Theta^{r}_{ix}(t-t',c)\bigg] \nonumber\\
&&+ \bigg[P_l(c+\frac{1}{n_0},x, t')\Theta^{d}_{lx}(t-t',c+\frac{1}{n_0})\nonumber\\
&&+\sum_{y\neq x}P(c-\frac{1}{n_0},x,t') \Theta^{d}_{ly}(t-t',c-\frac{1}{n_0})\lambda_{yx}\nonumber\\
&&-P_l(c,x,t')\Theta^{d}_{lx}(t-t',c)\nonumber\\
&&-\sum_{y\neq x}P_l(c,x,t')\Theta^{d}_{ly}(t-t',c)\lambda_{yx}\bigg]\bigg\}dt'.
\end{eqnarray}

Since
$\langle C_l(x,t)\rangle=\sum_{n_{xl}=0}^{n_0}\frac{n_{xl}}{n_0}P_l(c,x,t),$  then
\begin{eqnarray}
\langle\frac{\partial C_l(x,t)}{\partial t}\rangle&&=\int_0^1 c\frac{\partial P_l(c,x,t)}{\partial t}dc,
\end{eqnarray}
which can be written as
\begin{eqnarray}
\bigg\langle\frac{\partial C_l(x,t)}{\partial t}\bigg\rangle&&=\int_0^tdt'\int_0^1\bigg\{\bigg[\sum_{i:a_{ixl}\neq 0}  (c-a_{ixl}+a_{ixl})P_l(c-a_{ixl},x,t') \nonumber\\
&&\cdot\Theta^{r}_{ix}(t-t',c-a_{ixl})-cP_l(c,x,t') \Theta^{r}_{ix}(t-t',c)\bigg] \nonumber\\
&&+ (c+\frac{1}{n_0}-\frac{1}{n_0})P_l(c+\frac{1}{n_0},x, t')\Theta^{d}_{lx}(t-t',c+\frac{1}{n_0})\nonumber\\
&&-cP_l(c,x, t')\Theta^{d}_{lx}(t-t',c)\nonumber\\
&&+\sum_{y\neq x}\bigg[(c-\frac{1}{n_0}+\frac{1}{n_0})P_l(c-\frac{1}{n_0},x,t') \Theta^{d}_{ly}(t-t',c-\frac{1}{n_0})\lambda_{yx}\nonumber\\
&&-cP_l(c,x,t')\Theta^{d}_{ly}(t-t')\lambda_{yx})\bigg]\bigg\}dc.
\end{eqnarray}
We simplify the above equation, and get
\begin{eqnarray}
\frac{\partial \langle C_l(x,t)\rangle}{\partial t}&&=
\int_0^t\bigg[\sum_{i:a_{ixl}\neq 0}  a_{ixl}\langle\Theta^{r}_{ix}(t',t-t')\rangle\nonumber\\
&&-\frac{1}{n_0}\langle\Theta^{d}_{lx}(t',t-t')\rangle+\sum_{y\neq x}\frac{1}{n_0}\langle\Theta^{d}_{ly}(t',t-t')\rangle\lambda_{yx})\bigg]dt'.~~~~
\label{massactionrate}
\end{eqnarray}
Here, $\langle\Theta^{r}_{ix}(t',t-t')\rangle=\int_0^1 \Theta^{r}_{ix}(t-t',c)P_l(c,x,t')dc$, and $\langle\Theta^{d}_{lx}(t',t-t')\rangle=\int_0^1 \Theta^{d}_{lx}(t-t',c)P_l(c,x,t')dc.$
This equation is the mass action law for the reaction-diffusion system in one-dimensional finite lattice.
If there is no diffusion in the system, the above equation recovers the mass action law in Ref.\cite{AD2017}.
Note that although our model is performed in finite lattice, it also can be easily extended to infinite lattice where the transition distribution $\lambda_{yx}$ from $y$ to $x$ will be changed into infinite case.

\section{An example: monomolecular $A\leftrightarrow B$ reaction-diffusion system}
As an example, we shall consider the  monomolecular $A\leftrightarrow B$ reaction-diffusion system in which the A and B particles can move in one-dimensional finite lattice.
There are two reactions which make the change of the concentrations of A and B. The first reaction, we denote by 1, is the reaction from A to B, and the second we denote by 2  is the reaction from B to A.
Then the one-dimensional mass action rate (\ref{massactionrate}) becomes
\begin{eqnarray}
\frac{\partial < C_A(x,t)>}{\partial t}&&=\frac{1}{n_0}\int_{0}^{t}[-\langle\Theta_{1x}^{r}(t',t-t')\rangle+\langle\Theta_{2x}^{r}(t',t-t')\rangle-\langle\Theta_{Ax}^{d}(t',t-t')\rangle\nonumber\\
&&+\sum_{y\neq x}\int_{0}^{t}\langle\Theta_{Ay}^{d}(t',t-t')\rangle\lambda_{yx}] dt',
\label{masterA}
\end{eqnarray}
and
\begin{eqnarray}
\frac{\partial < C_B(x,t)>}{\partial t}&&=\frac{1}{n_0}\int_{0}^{t}[-\langle\Theta_{2x}^{r}(t',t-t')\rangle+\langle\Theta_{1x}^{r}(t',t-t')\rangle-\langle\Theta_{Bx}^{d}(t',t-t')\rangle\nonumber\\
&&+\sum_{y\neq x}\langle\Theta_{By}^{d}(t',t-t')\rangle\lambda_{yx}]dt',
\label{masterB}
\end{eqnarray}
where $C_A(x,t)=\frac{n_{xA}}{n_0}$ and $C_B(x,t)=\frac{n_{xB}}{n_0}$.

\subsection{Exponential case}
For exponent waiting time, that is, $p_{1x}(t)=\alpha_{rAx}e^{-\alpha_{rAx}t}$ and $p_{2x}(t)=\alpha_{rBx}e^{-\alpha_{rBx}t}$,
and $p_{dAx}(t)=\alpha_{dAx}e^{-\alpha_{dAx}t}$ and $p_{dBx}(t)=\alpha_{dBx}e^{-\alpha_{dBx}t}$, then we have
\begin{eqnarray}
\psi_{1x}^{r}(t,k)=n_{xA}\alpha_{rAx}\exp(-\alpha_{rAx}n_{xA}t)
\end{eqnarray}
\begin{eqnarray}
\psi_{2x}^{r}(t,k)=n_{xB}\alpha_{rBx}\exp(-\alpha_{rBx}n_{xB}t)
\end{eqnarray}
\begin{eqnarray}
\Psi_{1x}^{r}(t,k)=\exp(-\alpha_{rAx}n_{xA}t)
\end{eqnarray}
\begin{eqnarray}
\Psi_{2x}^{r}(t,k)=\exp(-\alpha_{rBx}n_{xB}t)
\end{eqnarray}
\begin{eqnarray}
\psi_{lx}^{d}(t,k)=n_{xA} \alpha_{dAx}\exp(-\alpha_{dAx}n_{xA}t)
\end{eqnarray}
\begin{eqnarray}
\psi_{2x}^{d}(t,k)=n_{xB} \alpha_{dBx}\exp(-\alpha_{dBx}n_{xB}t)
\end{eqnarray}
\begin{eqnarray}
\Psi_{1x}^{d}(t,k)=\exp(-\alpha_{dAx}n_{xA}t)
\end{eqnarray}
\begin{eqnarray}
\Psi_{2x}^{d}(t,k)=\exp(-\alpha_{dBx}n_{xB}t)
\end{eqnarray}
\begin{eqnarray}
\phi_{1x}^{r}(t,k)&=&n_{xA}\alpha_{rAx}\exp\{-\sum_{x=1,2,...,\upsilon}[\alpha_{rAx}n_{xA}+\alpha_{rBx}n_{xB}\nonumber\\
&&+\alpha_{dAx}n_{xA}+\alpha_{dBx}n_{xB}]t\}
\end{eqnarray}
\begin{eqnarray}
\phi_{2x}^{r}(t,k)&=&n_{xB}\alpha_{rBx}\exp\{-\sum_{x=1,2,...,\upsilon}[\alpha_{rAx}n_{xA}+\alpha_{rBx}n_{xB}\nonumber\\
&&+\alpha_{dAx}n_{xA}+\alpha_{dBx}n_{xB}]t\}
\end{eqnarray}
and
\begin{eqnarray}
\phi_{lx}^{d}(t,k)&=&n_{xA}\alpha_{dAx}\exp\{-\sum_{x=1,2,...,\upsilon}[\alpha_{rAx}n_{xA}+\alpha_{rBx}n_{xB}\nonumber\\
&&+\alpha_{dAx}n_{xA}+\alpha_{dBx}n_{xB}]t\}
\end{eqnarray}
\begin{eqnarray}
\phi_{2x}^{d}(t,k)&=&n_{xB}\alpha_{dBx}\exp\{-\sum_{x=1,2,...,\upsilon}[\alpha_{rAx}n_{xA}+\alpha_{rBx}n_{xB}\nonumber\\
&&+\alpha_{dAx}n_{xA}+\alpha_{dBx}n_{xB}]t\}
\end{eqnarray}
\begin{eqnarray}
\Phi(t,k)&=&\exp\{-\sum_{x=1,2,...,\upsilon}[\alpha_{rAx}n_{xA}+\alpha_{rBx}n_{xB}\nonumber\\
&&+\alpha_{dAx}n_{xA}+\alpha_{dBx}n_{xB}]t\},
\end{eqnarray}
and then
$$\Theta_{1x}^{r}(u,k)=n_{xA}\alpha_{rAx},$$
$$\Theta_{2x}^{r}(u,k)=n_{xB}\alpha_{rBx},$$
and
$$\Theta_{Ax}^{d}(u,k)=n_{xA}\alpha_{dAx},$$
$$\Theta_{Bx}^{d}(u,k)=n_{xB}\alpha_{dBx}.$$
Inverting them to time space yields
$$\Theta_{1x}^{r}(t,k)=n_{xA}\alpha_{rAx}\delta(t),$$
$$\Theta_{2x}^{r}(t,k)=n_{xB}\alpha_{rBx}\delta(t),$$
and
$$\Theta_{Ax}^{d}(t,k)=n_{xA}\alpha_{dAx}\delta(t),$$
\begin{equation}
\Theta_{Bx}^{d}(t,k)=n_{xB}\alpha_{dBx}\delta(t).
\end{equation}
Thus, we get
$$
\frac{1}{n_0}\int_{0}^{t}\langle\Theta_{1x}^{r}(t',t-t')\rangle dt'=\alpha_{rAx}\langle C_A(x,t)\rangle,$$
$$\frac{1}{n_0}\int_{0}^{t}\langle\Theta_{2x}^{r}(t',t-t')\rangle dt'=\alpha_{rBx}\langle C_B(x,t)\rangle$$
$$\frac{1}{n_0}\int_{0}^{t}\langle\Theta_{Ax}^{d}(t',t-t')\rangle dt'=\alpha_{dAx}\langle C_A(x,t)\rangle,$$
$$\frac{1}{n_0}\int_{0}^{t}\langle\Theta_{Ay}^{d}(t',t-t')\rangle dt'=\alpha_{dAy}\langle C_A(y,t)\rangle,$$
$$\frac{1}{n_0}\int_{0}^{t}\langle\Theta_{Bx}^{d}(t',t-t')\rangle dt'=\alpha_{dBx}\langle C_B(x,t)\rangle,$$
and
$$\frac{1}{n_0}\int_{0}^{t}\langle\Theta_{By}^{d}(t',t-t')\rangle dt'=\alpha_{dBy}\langle C_B(y,t)\rangle.$$

We substitute them into the mass action law
(\ref{masterA}) and (\ref{masterB}), and obtain
\begin{eqnarray}
\frac{\partial < C_A(x,t)>}{\partial t}&&=-\alpha_{rAx}<C_A(x,t))>\nonumber\\
&&+\alpha_{rBx}<C_B(x,t))>-\alpha_{dAx}<C_A(x,t))\nonumber\\
&&+\sum_{y\neq x} \alpha_{dAy}<C_A(y,t)>\lambda_{yx},
\end{eqnarray}
and
\begin{eqnarray}
\frac{\partial < C_B(x,t)>}{\partial t}&&=-\alpha_{rBx}<C_B(x,t))>\nonumber\\
&&+\alpha_{rAx}<C_A(x,t))>-\alpha_{dBx}<C_B(x,t))\nonumber\\
&&+\sum_{y\neq x} \alpha_{dBy}<C_B(y,t)>\lambda_{yx},
\end{eqnarray}
which are the classical mass action laws
for the $A\leftrightarrow B$ reaction-diffusion system.

\subsection{Power-law case}
In this case we consider the PDFs for waiting times obeying power-law distribution, that is,
$p_{1r}(t)\sim \frac{\tau_0^{\beta_{1r}}\beta_{1r}}{\Gamma(1-\beta_{1r})}\frac{1}{t^{1+\beta_{1r}}}, p_{2r}(t)\sim \frac{\tau_0^{\beta_{2r}}\beta_{2r}}{\Gamma(1-\beta_{2r})}\frac{1}{t^{1+\beta_{2r}}}$,
$p_{dA}(t)=\frac{\tau_0^{\beta_{dA}}\beta_{dA}}{\Gamma(1-\beta_{dA})}\frac{1}{t^{1+\beta_{dA}}}$ and $p_{dB}(t)=\frac{\tau_0^{\beta_{dB}}\beta_{dB}}{\Gamma(1-\beta_{dB})}\frac{1}{t^{1+\beta_{dB}}}$ where $0<\beta_{1r},\beta_{2r},\beta_{dA}, \beta_{dB}<1$.

Then we have

\begin{eqnarray}
\phi_{1x}^{r}(t,k)=n_{xA}\beta_{1r}\frac{\tau_0^{H}}{F}\frac{1}{t^{1+H}}
\end{eqnarray}
Taking the Laplace transform $t\rightarrow u$, we obtain
\begin{eqnarray}
\phi_{1x}^{r}(u,k)\sim \frac{n_{xA}\beta_{1r}}{H}[1-\frac{\Gamma(1-H)}{F}(\tau_0 u)^{H}].
\end{eqnarray}
Analogously, one can get
\begin{eqnarray}
\Phi(u,k)\sim \frac{\Gamma(1-H)}{F}(\tau_0 u)^{H-1},
\end{eqnarray}
Therefore,
\begin{eqnarray}
\Theta_{1x}^r(u)\sim n_{xA}\beta_{1r}Mu^{1-H}.
\end{eqnarray}
Taking the Laplace transform of $\frac{1}{n_0}\int_{0}^{t}\langle \Theta_{1x}^{r}(t',t-t')\rangle $, we find
\begin{eqnarray}
\textbf{L}\bigg\{\frac{1}{n_0}\int_{0}^{t}\langle \Theta_{1x}^{r}(t',t-t')\rangle dt'\bigg\}
&&=\frac{1}{n_0}\int_{0}^{1}p(c,x,u)\Theta_{1x}^{r}(u,c)dc\nonumber\\
&&=\int_{0}^{1}p(c,x,u)\frac{n_{xA}}{n_0}Mu^{1-H}\beta_{1r}dc\nonumber\\
&&=\beta_{1r}Mu^{1-H}\langle C_A(x,u)\rangle,
\end{eqnarray}
whose inverse Laplace transform is
\begin{equation}
\frac{1}{n_0}\int_{0}^{t}\langle\Theta_{1x}^{r}(t',t-t')\rangle dt'=\beta_{1r}M\mathcal{D}_t^{1-H}\langle C_A(x,t)\rangle,
\label{powertheta1}
\end{equation}
\begin{equation}
\frac{1}{n_0}\int_{0}^{t}\langle\Theta_{2x}^{r}(t',t-t')\rangle dt'=\beta_{2r}M\mathcal{D}_t^{1-H}\langle C_B(x,t)\rangle,
\label{powertheta2}
\end{equation}
\begin{equation}
\frac{1}{n_0}\int_{0}^{t}\langle\Theta_{Ax}^{d}(t',t-t')\rangle dt'=\beta_{dA}M\mathcal{D}_t^{1-H}\langle C_A(x,t)\rangle,
\label{powertheta3}
\end{equation}
\begin{equation}
\frac{1}{n_0}\int_{0}^{t}\langle\Theta_{Bx}^{r}(t',t-t')\rangle dt'=\beta_{dB}M\mathcal{D}_t^{1-H}\langle C_B(x,t)\rangle,
\label{powertheta4}
\end{equation}
where $M=\frac{1}{H}\frac{1}{\tau_0^H}\frac{F}{\Gamma(1-H)}$,
with $F=\Gamma^{n_A}(1-\beta_{1r})\Gamma^{n_B}(1-\beta_{2r})\Gamma^{n_A}(1-\beta_{dA})\Gamma^{n_B}(1-\beta_{dB}),
$  and $H=\beta_{1r}n_A+\beta_{2r}n_B+\beta_{dA}n_A+\beta_{dB}n_B$, $0<H<1$.
We substitute them into the mass action laws
(\ref{masterA}) and (\ref{masterB}), and obtain
\begin{eqnarray}
\frac{\partial < C_A(x,t)>}{\partial t}&&=-\beta_{1r}M\mathcal{D}_t^{1-H}\langle C_A(x,t)\rangle\nonumber\\
&&+\beta_{2r}M\mathcal{D}_t^{1-H}\langle C_B(x,t)\rangle-\beta_{dA}M\mathcal{D}_t^{1-H}\langle C_A(x,t)\rangle\nonumber\\
&&+\sum_{y\neq x} \beta_{dA}M\mathcal{D}_t^{1-H}\langle C_A(y,t)\rangle\lambda_{yx},
\label{amal}
\end{eqnarray}
and
\begin{eqnarray}
\frac{\partial < C_B(x,t)>}{\partial t}&&=-\beta_{2r}M\mathcal{D}_t^{1-H}\langle C_B(x,t)\rangle\nonumber\\
&&+\beta_{1r}M\mathcal{D}_t^{1-H}\langle C_A(x,t)\rangle-\beta_{dB}M\mathcal{D}_t^{1-H}\langle C_B(x,t)\rangle\nonumber\\
&&+\sum_{y\neq x}  \beta_{dB}M\mathcal{D}_t^{1-H}\langle C_B(y,t)\rangle\lambda_{yx}.
\label{bmal}
\end{eqnarray}
Here, $\mathcal{D}_t^{1-H}$ is the Riemann-Liouville fractional
differential operator. This means that the concentration depends on the history evolution for A and B particles.

\section{Generalized Gillespie algorithm}
Based on the chemical continuous time random walks under anomalous diffusion model, we can get the generalized Gillespie algorithm as follows.

(i) In the initial state $n_0$ reactants are
randomly placed at $\upsilon$ places by uniform distribution.

(ii) The random
waiting time $\tau_{ix}$ as the internal clock to react for
each reaction $i$ at $x$  is chosen from a series of values distributed
according to $p_{ix}(t)$ and the random
waiting time $\tau_{dlx}$ as the internal clock to make next jump for
each reactant $l$ at $x$ is chosen from a series of values distributed
according to $p_{dlx}(t)$ [23-25]. If $p_{ix}(t)$ is exponential distribution $p_{ix} (t)=\alpha_{ix}e^{-\alpha_{ix}t}$,  then we get $\tau_{ix}=-\frac{ln u}{\alpha_{ix}}$, where $u$ is a random variate drawn from the uniform density
on the interval $[0,1]$. If the $p_{ix}(t)$ is power-law distribution $p_{ix} (t)=\alpha_{ix}\tau_0^{\alpha_{ix}}t^{-(1+\alpha_{ix})}$,  then we get
$\tau_{ix}=\tau_0 u^{-\frac{1}{\alpha_{ix}}}$. The generation of $\tau_{dlx}$ is as same as that of $\tau_{ix}.$

(iii) Find the minimum clock time $\min\{\tau_{ix},\tau_{dlx}\}(i=1,2,...,m; l=1,2,...,\gamma; x=1,2,...,\upsilon)$. If the minimum waiting time is for one reaction $i$ at $x$ then we change the system according to this reaction \cite{MR2018}, or the minimum waiting time is for one diffusion process for a reactant $l$ at $x$, then
we let $l$ diffuse from $x$ to one place $y$ with the
probability $\lambda_{xy}$, after which the internal clocks for reactions and diffusion
are all reset by the distributions and repeat the procedure.

Note that without the consideration of the effect of diffusion, this algorithm is equivalent to the non-Markovian Gillespie algorithm for reactive system where the reaction $i$ is chosen by the instantaneous probability
\begin{eqnarray}\label{rate}
\gamma(t,k)=\lim_{\Delta t\rightarrow 0}\frac{\textbf{P}(t<\tau<t+\Delta t|\tau>t)}{\Delta t},
\end{eqnarray}
after the time $\Delta t$ for the next event drawn in
\cite{MR2018}.
Notice that $\gamma(t,k)$ is just the rate of escape (hazard function) $\gamma(t,k)$ which can be written as
 \cite{CM1965, FIZ2013}
\begin{eqnarray}
\gamma(t,k)=\frac{\phi(t,k)}{\Phi(t,k)},
\label{gamma}
\end{eqnarray}
where
\begin{eqnarray}
\Phi(t,k)&=&\Pi_{x=1,2,...,\upsilon;i=1,2,...,m}\Psi_{ix}^{r}(t,k),
\end{eqnarray}
is the survival probability and
\begin{eqnarray}
\phi(t,k)=-\frac{\partial \Phi(t,k)}{\partial t}=\sum_{x=1}^{\upsilon}\sum_{i=1}^{m}\phi_{ix}^{r}(t,k).
\end{eqnarray}
From the above two equations one has
\begin{eqnarray}
\Phi(t,k)=e^{-\int_{0}^{t}\gamma(\tau,k)d\tau},
\end{eqnarray}
and
\begin{eqnarray}
\phi(t,k)=\gamma(t,k)e^{-\int_{0}^{t}\gamma(\tau,k)d\tau},
\end{eqnarray}
This means that one can get $\Phi(t,k)$ and $\phi(t,k)$ by using the rate $\gamma(t,k)$ and the relationship $\phi(t,k)=-\frac{\partial \Phi(t,k)}{\partial t}$, in reverse, from the waiting time PDF $\phi(t,k)$ and the relationship, one can also get  $\Phi(t,k)$, which means that the two approaches are equivalent in the case without considering diffusion in the reaction-diffusion system.

\section{Comparison with the mass action law derived from continuous time random walk (CTRW) models.}
In this section we shall compare our mass action law with that derived from continuous time random walk (CTRW) for the state A to B and backward under anomalous transport by Campos e.t. in \cite{CFM2008}.  The corresponding master equations in \cite{CFM2008} is as following
\begin{eqnarray}\label{amasterdiffusion}
\frac{dP_1(x,t)}{dt}&&=\int_{-\infty}^{+\infty}\int_{0}^{t}P_{1}(x',t')\Theta_{1j}(t-t')\lambda_1(x-x')dt'dx'\nonumber\\
&&+\int_{0}^{t}P_{2}(x,t)\Theta_{2r}(t-t')dt'-\int_{0}^{t}P_{1}(x,t')\Theta_{1j}(t-t')dt'-\int_{0}^{t}P_{1}(x,t')\Theta_{1r}(t-t')dt',
\end{eqnarray}
\begin{eqnarray}\label{bmasterdiffusion}
\frac{dP_2(x,t)}{dt}&&=\int_{-\infty}^{+\infty}\int_{0}^{t}P_{2}(x',t')\Theta_{2j}(t-t')\lambda_2(x-x')dt'dx'\nonumber\\
&&+\int_{0}^{t}P_{1}(x,t')\Theta_{1r}(t-t')dt'-\int_{0}^{t}P_{2}(x,t')\Theta_{2j}(t-t')dt'-\int_{0}^{t}P_{2}(x,t')\Theta_{2r}(t-t')dt'.
\end{eqnarray}
Here, $P_{1,2}(x,t)=\frac{N_{1,2}(x,t)}{N}$ respectively denote the PDFs for a particle being in state $A$ and $B$ at point $x$ at time $t$  where  $N_{1,2}(x,t)$  are the
the densities of $A$ and $B$, and $N$ is the sum of $N_{1}(x,t)$ and $N_2(x,t)$ over the whole space. One can see that $\int_{-\infty}^{+\infty}(P_1(x,t)+P_2(x,t))dx=1$.
$\lambda_{1,2}(\Delta x)$ denote the jump length PDFs for  $A$ particle and $B$ particle, respectively. $\psi_{1j,2j}(t)$ denote the PDFs of random diffusion waiting times, $\psi_{1r,2r}(t)$ denote the PDFs of random  reaction waiting time  of $A\to B$ and $B\to A$,  $\Psi_{1j,2j}(t)=1-\int_{0}^{t}\psi_{1j}(\tau)d\tau$ and $\Psi_{1r,2r}(t)=1-\int_{0}^{t}\psi_{2j}(\tau)d\tau$ are the corresponding survival probabilities for $\psi_{1j,2j}(t)$ and $\psi_{1j,2j}(t)$. Finally,  $\Theta_{1j}(u)=\frac{\textit{L}(\Psi_{1r}(t)\psi_{1j}(t))}{\textit{L}(\Psi_{1r}(t)\Psi_{1j}(t))}$
and $\Theta_{1r}(u)=\frac{\textit{L}(\Psi_{1j}(t)\psi_{1r}(t))}{\textit{L}(\Psi_{1j}(t)\Psi_{1r}(t))}$ are the kernels.

When the random jump length is Gaussian satisfying
\begin{eqnarray}\label{gjumplength}
\lambda(k)\sim
1-\frac{\sigma^{2}k^{2}}{2}
\end{eqnarray}
and random waiting times for reaction and diffusion are both power-law distributed satisfying
\begin{eqnarray}\label{powerlawdistributedr}
\psi_{1r,2r}(t)=\beta_{1r,2r}\tau ^{\beta_{1r,2r}}\frac{1}{t^{1+\beta_{1r,2r}}}
\end{eqnarray}
and
\begin{eqnarray}\label{powerlawdistributedj}
\psi_{1j,2j}(t)=\beta_{1j,2j}\tau ^{\beta_{1j,2j}}\frac{1}{t^{1+\beta_{1j,2j}}},
\end{eqnarray}
where  $\sigma^{2}$ is
the jump length variance,
one gets  $\Theta_{hj}(u)\sim\frac{\beta_{hj}}{\beta_{hr}+\beta_{hj}}\frac{u^{1-\beta_{hr}-\beta_{hj}}}{\tau^{\beta_{hr}+\beta_{hj}}\Gamma(1-\beta_{hr}-\beta_{hj})}$
and
$\Theta_{hr}(u)\sim\frac{\beta_{hr}}{\beta_{hr}+\beta_{hj}}\frac{u^{1-\beta_{hj}-\beta_{hr}}}{\tau^{\beta_{hj}+\beta_{hr}}\Gamma(1-\beta_{hj}-\beta_{hr})}$ for $0<\beta_{1r}+\beta_{1j},\beta_{2r}+\beta_{2j}<1$.

Taking the Fourier-Laplace transform of the generalized master equations (\ref{amasterdiffusion}) and (\ref{bmasterdiffusion}) yields
\begin{eqnarray}\label{masterkuA}
&&u P_1(k,u)-P_1(k,0)=P_1(k,u)\Theta_{1j}(u)(-\frac{\sigma^{2}k^{2}}{2})\nonumber\\
&&+P_2(k,u)\Theta_{2r}(u)-P_1(k,u)\Theta_{1r}(u),
\end{eqnarray}
and
\begin{eqnarray}\label{masterkuB}
&&u P_2(k,u)-P_2(k,0)=P_2(k,u)\Theta_{2j}(u)(-\frac{\sigma^{2}k^{2}}{2})\nonumber\\
&&+P_1(k,u)\Theta_{1r}(u)-P_2(k,u)\Theta_{2r}(u).
\end{eqnarray}
By substituting $\Theta_{hj}(u)$ and $\Theta_{hr}(u)$ into (\ref{masterkuA}) and (\ref{masterkuB}) and taking the  inverse Fourier-Laplace transform, we finally obtain
the generalized fractional reaction-diffusion equations
\begin{eqnarray}\label{agfrde}
\frac{dP_1(x,t)}{dt}&=&C_{11}\frac{\partial^2 D_t^{1-\beta_{1r}-\beta_{1j}} P_1(x,t)}{\partial x^2}+C_{2}D_t^{1-\beta_{2j}-\beta_{2r}}P_{2}(x,t)-C_{3} D_t^{1-\beta_{1j}-\beta{1r}}P_{1}(x,t),
\end{eqnarray}
\begin{eqnarray}\label{bgfrde}
\frac{dP_2(x,t)}{dt}&=&C_{21}\frac{\partial^2 D_t^{1-\beta_{2r}-\beta_{2j}} P_2(x,t)}{\partial x^2}-C_{2}D_t^{1-\beta_{2j}-\beta_{2r}}P_{2}(x,t)+C_{3} D_t^{1-\beta_{1j}-\beta{1r}}P_{1}(x,t),
\end{eqnarray}
with $$C_{11}=\frac{\sigma^2}{2}\frac{\beta_{1j}}{\beta_{1r}+\beta_{1j}}\frac{1}{\tau^{\beta_{1r}+\beta_{1j}}\Gamma(1-\beta_{1r}-\beta_{1j})},$$
$$C_{21}=\frac{\sigma^2}{2}\frac{\beta_{2j}}{\beta_{2r}+\beta_{2j}}\frac{1}{\tau^{\beta_{2r}+\beta_{2j}}\Gamma(1-\beta_{2r}-\beta_{2j})},$$
$$C_{2}=\frac{\beta_{2r}}{\beta_{2r}+\beta_{2j}}\frac{1}{\tau^{\beta_{2r}+\beta_{2j}}\Gamma(1-\beta_{2r}-\beta_{2j})},$$
$$C_{3}=\frac{\beta_{1r}}{\beta_{1r}+\beta_{1j}}\frac{1}{\tau^{\beta_{1r}+\beta_{1j}}\Gamma(1-\beta_{1r}-\beta_{1j})}.$$
When we assume that there is only one particle in the reaction-diffusion system,  the jump length is Gaussian whose PDF obeys (\ref{gjumplength}) and the PDFs of reaction and diffusion waiting times
is space-independent satisfying (\ref{powerlawdistributedr}) and (\ref{powerlawdistributedj}), we find that our mass action law (\ref{amal}) and (\ref{bmal}) can recover (\ref{agfrde}) and (\ref{bgfrde}).

\section{Comparison with the chemical continuous time random walk (CTRW) models.}

In Ref.\cite{AD2017} it defines an inhomogeneous continuous time random walk in particle number space, from which
a generalized chemical master equation is derived as following
\begin{eqnarray}
\frac{\partial P(n,t)}{\partial t}&&=\sum_i \int_0^t dt'(\prod_j \textit{E}_j^{-s_{ij}}-1)P(n,t')M_i(t-t',n),
\label{ADmaster}
\end{eqnarray}
where the step operator $\textit{E}_j^{z}$
acts on a function $f(n)$ by incrementing
the particle number $n_j$ of species $S_j$. $M_i$ is the memory function.
This equation is the special case of the mass action law (\ref{massactionrate}) without diffusion in the system.

Based on the derived master equation (\ref{ADmaster}), the authors determine the modified chemical rate laws
\begin{eqnarray}
\frac{\partial \langle C\rangle}{\partial t}&&=\sum_i s_i\int_0^t dt'M_i^C(t-t',C),
\label{master_rate}
\end{eqnarray}
where $M_i^C(t-t',C)=\frac{M_i^C(t-t',n_0 C)}{n_0}$.
Note that when the diffusion is ignored, namely, $\lambda (y-x)=0$ in our rate laws (\ref{massactionrate}) we can recover (\ref{master_rate}).

Furthermore, a generalization of the Gillespie
algorithm is proposed for the reaction system in \cite{AD2017}. Comparing with their algorithm, we find that  their algorithm is specially for exponential reactive waiting times, while our algorithm is for arbitrary distributed and more general.  Beside, we also  take account of the diffusion for generalization while it is not considered in \cite{AD2017}. But they further consider the global delay that is added to inter-reaction waiting time.

\section {Conclusion}
To sum up, in order to describe the coupled non-Poissonian reaction and anomalous diffusion process in heterogeneous system, where the reactive and diffusive waiting times are both arbitrary distributions, we develop a chemical continuous time random walk under anomalous diffusion model. We firstly use the renewal process to describe the numbers of reaction and diffusion renewals,  and also define the distribution of the  minimum renewal waiting times. Based on the renewal process we obtain the  generalized chemical diffusion master equation (19) for the probability of the numbers of reactants, and then derive the corresponding  mass action law (26) for the average concentration of the reactants using the CDME. Moreover, we  generalize the Gillespie algorithm to fit for the non-Markovian reaction-diffusion system. In the end, we take the  $A\leftrightarrow B$ reaction system in which the A and B particles for exponential and power-law waiting times respectively. We recover the mass law rate in the exponential waiting time case, and show the strong fractional memory effect of the concentration of the reactants on the history of the concentration in the power-law  case.

\vspace*{2mm}

\section{Acknowledgements}
Much of the work in this paper was finished when Hong Zhang visited the University of Manchester. We appreciate Prof. Sergei Fedotov for his valuable suggestions.
This work was supported by the Natural Science Foundation of Sichuan Province (Grant
No. 2022NSFSC1752).

\section{Conflict of interest statement}
The authors declare that there are no conflict of interests,  we do not have any possible conflicts of interest.

\section{Data availability}
Date sharing is not applicable to this paper as no new data were created or analyzed in this study.

\end{document}